 \documentclass{aip-cp}  
\usepackage[numbers]{natbib}
\usepackage{rotating}
\usepackage{graphicx}

\usepackage{epsfig}
\usepackage{psfrag}

\newcommand{\be}{\begin{equation}}
\newcommand{\ee}{\end{equation}}
\newcommand{\bea}{\begin{eqnarray}}
\newcommand{\eea}{\end{eqnarray}}
\newcommand{\bno}{\begin{eqnarray*}}
\newcommand{\eno}{\end{eqnarray*}}

\newcommand{\bl}{\begin{large}}
\newcommand{\el}{\end{large}}
\newcommand{\bla}{\begin{Large}}
\newcommand{\ela}{\end{Large}}

\newcommand {\sla} {\slash \hspace{-0.22cm}}

\newcommand{\hsp}{\hspace{0.70cm}}  

\begin{document}

\title{Pion in the Medium with a Light-Front Model} 
\author[aff1]{J.~P.~B.~C.~de~Melo\corref{cor1}}
\author[aff1]{Kazuo Tsushima}
\eaddress{kazuo.tsushima@cruzeirodosul.edu.br}
\author[aff2]{Tobias Frederico}
\eaddress{tobias@fis.ita}
\affil[aff1]{Laborat\'orio de F\'\i sica Te\'orica e Computacional\\
Universidade Cruzeiro do Sul\\
01506-000, S\~ao Paulo, SP, Brazil}
\affil[aff2]{Instituto Tecnol\'ogico da Aeron\'autica/DCTA\\
12228-900, S\~ao Jos\'e dos Campos, SP, Brazil}
\corresp[cor1]{Corresponding author: 
joao.mello@cruzeirodosul.edu.br } 
\maketitle

\begin{abstract}
The pion properties in symmetric nuclear matter are investigated with 
the Quark-Meson Coupling (QMC) Model plus the light-front constituent quark model~(LFCQM). 
The LFCQM has been quite successful in describing the properties of  
pseudoscalar mesons in vacuum, such as the electromagnetic elastic form factors, 
electromagnetic radii, and decay constants. 
 We study the pion properties in 
symmetric nuclear matter with the in-medium input  
recalculated through the QMC model, which provides 
the in-medium modification of the LFCQM.
\end{abstract}

\section{INTRODUCTION}

\hsp 
A fundamental task in nuclear and particles physics is to understand the structure of 
hadronic systems in term of the quarks and gluons, 
where their interactions are described by the strong interaction of quantum 
chromodynamics~(QCD)~\cite{Muller1999,Hatsuda2010,Bakker2014}.  
Many experiments concerning the hadron properties are planned in some laboratories, 
among them, JLab~(see Ref.~\cite{Dudek2012} for details). 
However, another very important and interesting aspect with respect to hadronic physics  
is their properties in the nuclear medium. This includes the context of nuclear physics, 
i.e, NN interaction in a nucleus, neutron stars and particle properties  
in heavy ions collisions.
The main questions here is, 
``How the hadron properties change in the dense nuclear 
medium?'', and ``What is the effect of the nuclear medium  
on the QCD structure of hadrons?'' 

To answer these questions, we study here the pion properties in symmetric nuclear matter. 
Our approach is to use the quark-meson-coupling~(QMC)~model~
\cite{Guichon1988,Guichon1996,Saito2007} 
plus the light-front approach~\cite{deMelo2002,deMelo2014}. 
Many studies of the pion properties in the nuclear medium exist in the 
literature and readers are asked to consult e.g. Ref.~\cite{Hatsuda1985}.

In 1949, Dirac proposed three possible forms of relativistic dynamics~\cite{Dirac1949},
namely, instant form, point form and front form, and the last one  
is used in this work.

The use of the light-front form, instead of the instant form, has some advantages  
as follows.
Although the Fock state has infinite numbers of particles in general, 
only the valence component is necessary to calculate the electroweak 
properties of the hadronic systems~\cite{Brodsky1998}.
However in the light-front approach, it is possible to take into account
in the light-front wave function the  
higher Fock state components, 
which can be written in terms the lower ones~\cite{Brodsky1998,Pauli1998,Pauli1999}. 
Because of that, 
the light-front approach is an ideal framework to describe the  
hadronic bound states in terms 
of the valence component wave function, or in a picture 
of the constituent (quark) degrees of freedom. 
Thus, it can treat unambiguously  the parton (quark) content of the meson  
and baryon wave functions. 
Another important advantages are that the 
vacuum for the free Hamiltonian is trivial, and 
the light-front Hamiltonian is Lorentz invariant~\cite{Brodsky1998,Pauli1998}.
After the integration over the light-front energy~$(k^-=k^0-k^3)$  
of a given  Bethe-Salpeter amplitude, the hadronic valence wave function 
can be derived~\cite{deMelo1999,deMelo2002}.

\section{The Model} 

\hsp
For describing the nuclear matter, we use the QMC model 
developed in Ref.~\cite{Guichon1988}. 
(A similar approach using a confining potential was developed in Ref.~\cite{Frederico1989}.) 
QMC describes the nuclear matter based on  
the quark degrees of freedom. It has been successfully applied  
for studying the properties of finite nuclei~\cite{Saito1996}, and the hadronic 
properties in dense nuclear medium~(see Ref.~\cite{Saito2007},~for more details).

The effective Langrangian density for symmetric nuclear 
matter is given by~\cite{Saito2007},
\begin{equation}
{\cal L} = {\bar \psi} [i\gamma \cdot 
\partial -m_N^*({\hat \sigma}) -g_\omega {\hat \omega}^\mu \gamma_\mu ]
\psi
+ {\cal L}_\textrm{meson}, 
\label{lag1}
\end{equation}
where, $\psi$, $\hat{\sigma}$, and $\hat{\omega}$ are respectively the 
nucleon, Lorentz scalar-isoscalar, and Lorentz vector-isoscalar field operators 
with 
\begin{equation}
m_N^*({\hat \sigma}) = m_N - 
g_\sigma({\hat \sigma}) {\hat \sigma}.
\label{efnmas}
\end{equation}

The density ($\sigma$-field) dependent $\sigma$-N 
coupling constant in nuclear matter,~$g_{\sigma}(\hat{\sigma})$, 
is defined by Eq.~(\ref{efnmas}), and
$g_{\omega}$ is the $\omega$-N coupling constant. 
The meson Lagrangian density ${\cal L}_\textrm{meson}$ in Eq.~(\ref{efnmas}) is given by 
\begin{equation}
{\cal L}_\mathrm{meson} = \frac{1}{2} 
(\partial_\mu {\hat \sigma} \partial^\mu 
{\hat \sigma} - m_\sigma^2 {\hat \sigma}^2)
- \frac{1}{2} \partial_\mu {\hat \omega}_\nu (\partial^\mu {\hat \omega}^\nu - 
\partial^\nu {\hat \omega}^\mu)
+ \frac{1}{2} m_\omega^2 {\hat \omega}^\mu {\hat \omega}_\mu \ .
\label{mlag1}
\end{equation}
In the above the Lorentz vector-isovector dependence is ignored,  
because we considered the symmetric nuclear matter within the 
Hartree approximation~\cite{deMelo2014}.

We work in the nuclear matter rest frame hereafter.
The Dirac equations for the up ($u$) and down ($d$) quarks 
are solved self-consistently with the same mean values of the  
$\sigma$ and $\omega$ fields, which also act on the nucleon and describe the properties 
of nuclear matter~\cite{Saito1996,Saito2007}:  
\begin{eqnarray}
\left[ i \gamma \cdot \partial_x -
(m_q - V^q_\sigma)
\mp \gamma^0
\left( V^q_\omega +
\frac{1}{2} V^q_\rho
\right) \right]
\left( \begin{array}{c} \psi_u(x)  \nonumber \\
\psi_{\bar{u}}(x) \\ \end{array} \right) &=& 0,
\label{diracu}\\
\left[ i \gamma \cdot \partial_x -
(m_q - V^q_\sigma)
\mp \gamma^0
\left( V^q_\omega -
\frac{1}{2} V^q_\rho
\right) \right]
\left( \begin{array}{c} \psi_d(x)  \\
\psi_{\bar{d}}(x) \\ \end{array} \right) &=& 0,
\label{diracd}
\label{diracQ}
\end{eqnarray}
$SU(2)$ symmetry is assumed in the above for the quarks~$u$ and $d$. 
We define, $m^{*}_{q}\equiv m_q- V^q_{\sigma}=m^*_u=m^*_d$.  Also, in 
symmetric nuclear matter, the $\rho$-meson mean field potential,  
$V^q_{\rho}=0$, is dropped. The other mean-field potentials 
are defined by, $V^q_\sigma \equiv g^q_\sigma \sigma = g^q_\sigma <\sigma>$ and
$V^q_\omega \equiv g^q_\omega \omega = g^q_\omega\, \delta^{\mu,0} <\omega^\mu>$,
with $g^q_\sigma$ and $g^q_\omega$ the corresponding quark-meson coupling constants.

The baryon density ($\rho$) via $k_F$ the Fermi momentum, 
scalar density ($\rho_s$), and the 
total energy per nucleon ($E^\mathrm{tot}/A$) are given by,
\begin{eqnarray}
\rho &=& \frac{4}{(2\pi)^3}\int d\vec{k}\ \theta (k_F - |\vec{k}|)
= \frac{2 k_F^3}{3\pi^2}, \nonumber 
\label{rhoB} \\
\rho_s &=& \frac{4}{(2\pi)^3}\int d\vec{k} \ \theta (k_F - |\vec{k}|)
\frac{m_N^*(\sigma)}{\sqrt{m_N^{* 2}(\sigma)+\vec{k}^2}},
\label{rhos}\\
E^\mathrm{tot}/A&=&\frac{4}{(2\pi)^3 \rho}\int d\vec{k} \
\theta (k_F - |\vec{k}|) \sqrt{m_N^{* 2}(\sigma)+
\vec{k}^2}+\frac{m_\sigma^2 {\sigma}^2}{2 \rho}+
\frac{g_\omega^2\rho}{2m_\omega^2}, 
\label{toten}
\end{eqnarray}
where the nuclear matter saturation properties, namely, the binding energy per nucleon  
of~$15.7$~MeV~at the saturation density $\rho_0$ ($\rho_0=0.15~$fm$^{-3}$) are known, 
and from these the coupling constants~$g_{\sigma}$~and~$g_{\omega}$ are determined.

\begin{figure*}[t]
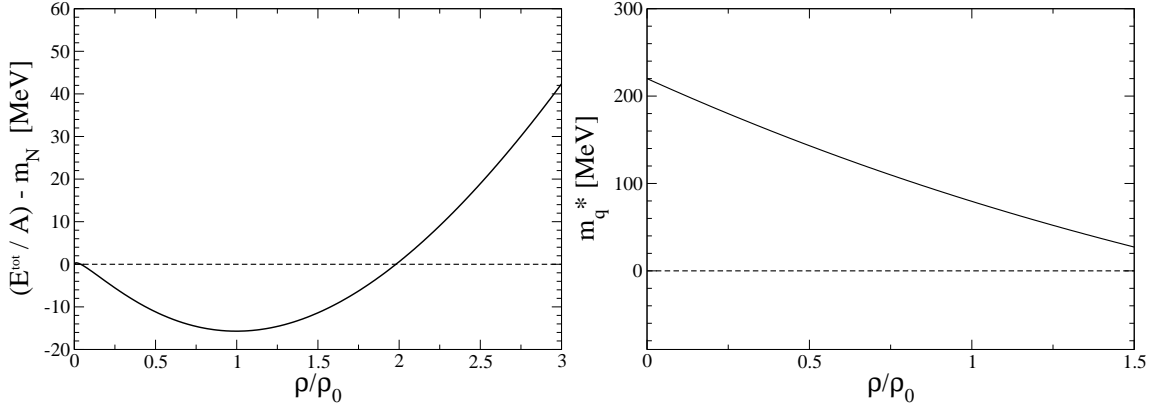

\begin{center}
\includegraphics[scale=0.30]{EnergyDen.eps}
\includegraphics[scale=0.30]{mqstar2.eps}
\caption{Left~(a)~~Negative of the binding energy per nucleon ($E^\mathrm{tot}/A - m_N$) for symmetric
nuclear matter calculated with the vacuum up and down quark masses, $m_q = 220$ MeV.
At the saturation point $\rho_0 = 0.15$ fm$^{-3}$, the value is fitted to $-15.7$~MeV.
Right~(b)~~Effective constituent mass for the up and down quarks in  
symmetric nuclear matter, $m^*_q = m^*_u = m^*_d$.
\label{Fig1ab}
}
\end{center}
\end{figure*}

In order to calculate the pion properties in symmetric nuclear matter, 
we use the light-front model of Ref.~\cite{deMelo2002}. 
That model reproduces quite well the pion experimental data, i,e., 
pion electromagnetic form factor, electromagnetic radius and also 
the weak decay constant. Then, in this work the 
light-front constituent quark model (LFQCM)  and QMC, 
are combined to study the pion properties in  
symmetric nuclear matter.

The pion properties in symmetric nuclear matter is calculated with the 
effective Lagrangian density with a pseudoscalar coupling of the quarks 
to the pion~\cite{Frederico1992}, used before for the vacuum case~\cite{deMelo2014},   
\begin{equation}
{{\cal L}_I= - i g^* \vec\Phi \cdot \overline q \gamma^5 
\vec \tau q \, \Lambda^*.}  
\label{lain}
\end{equation}
Here,  $g^*$  is the coupling constant, and $\Lambda^{*}$ is the vertex function in the medium.  
The coupling constant is given by the Goldberger-Treiman relation,~$g*=m^*_q/f^*_{\pi}$, 
using the quantities in symmetric nuclear matter (with the asterisks). 
 
The electromagnetic current in symmetric nuclear matter is obtained in the impulse 
approximation, represented by the
Feymman triangle diagram~\cite{deMelo2002},  
 \begin{eqnarray}
\hspace*{-1mm}
j^\mu & = &  -i\, 2 e \frac{m_q^{*2}}{f^{*2}_\pi} N_c \! 
\int\! \frac{d^4k'}{(2\pi)^4}\, \mathrm{Tr} \left[ 
S^*(k')\gamma^5 S^*(k'-P^{\prime})   
\gamma^\mu S^*(k'-P) 
\gamma^5   \right]
\Lambda^*(k',P^{\prime}) \Lambda^*(k',P)~,
\label{jmu}
\end{eqnarray}
where the factor 2 comes from the isospin algebra.

The in-medium modifications can be implemented with the model 
of Ref.~\cite{deMelo2002}. The  in-medium quark propagators are given by,
\begin{eqnarray}
\displaystyle S^*(p+V)=
\frac{1}{ \sla{p} + \sla{V}-m_q^*+ i\epsilon}.
\end{eqnarray}
In the symmetric nuclear matter, the quark properties are modified by the 
Lorentz-scalar-isoscalar and the Lorentz-vector-isoscalar mean field potentials. 
In mean field approximation, the modifications
enter as the shift in the quark and anti-quark momentum, 
$p^\mu ~\rightarrow~ p^\mu + V^\mu~=~p^\mu + \delta^\mu_0 V^0 $ and 
$p^\mu + V^\mu~=~p^\mu \pm \delta^\mu_0 V^q_\omega $, for the quark~(+) and 
anti-quark~(-), in the case of the vector potential. For the Lorentz scalar part, 
we have a modification in the quark mass as~$m_q \rightarrow m^*_q= m + V_s~(=m_q-V^q_\sigma)$.

Also, the in-medium vertex function is given by~\cite{deMelo2002,deMelo2014}, 
\begin{equation}
\Lambda^*(k+V,P)=
\frac{C^*}{((k+V)^2-m^2_{R} + i\epsilon)}+
\frac{C^*}{((P-k-V)^2-m^2_{R}+ i\epsilon)}.
\label{vertex}
\end{equation}
The main motivation to choose the above symmetric regulator is that, this vertex function is 
symmetric by the exchange of the momentum between the two fermions, and we 
have a symmetric light-front valence wave function~\cite{deMelo2014,deMelo2002}.

In the present case the effect of the vector potentials in the loop integral 
associated with the Feynman diagram is cancel out, because of our choice of 
the  pion vertex. Then, only the mass 
shift of the quarks is relevant for the loop integral. 
(See Ref.~\cite{deMelo2014} for details on this point.)
In the vertex function Eq.~(\ref{vertex}), the parameter $C^*$  
is assumed to be the same as that in vacuum,  
because it is associated with the short-range scale of the wave function of the pion.

In the present work, we use the Breit-frame and the Drell-Yan condition~($q^+=0$),
where the  momentum transfer is 
given by~$q^\mu=(+,-,\perp)=(0,0,q/2)=(P'-P)^\mu$, $q^+=-q^-=0$,~$q_x=-q/2$ 
and~$q^2 = q^+q^--(\vec q_\perp)^2 \equiv -Q^2$. 
The bound state mass is~$m_B= \sqrt{P^{+2}+q^2/4}$ with $P^{+}=P^{\prime+}$.
In order to calculate the covariant form factor, we use the 
following expression
\begin{equation}
   j^\mu = e (P^{\mu}+P^{\prime \mu}) F^*_\pi (q^2).
\label{full}
\end{equation}

The pion elastic electromagnetic form factor has two contributions in 
the light-front approach~\cite{deMelo1999,deMelo2002,Yabuzaki2015}, i.e., the 
valence and the non-valence contributions~(see the Fig.~2):
\begin{eqnarray}
F^*_\pi(q^2)=F^{*(I)}_\pi(q^2)+F^{*(II)}_\pi(q^2).
\label{ffactor}
\end{eqnarray}

The Bethe-Salpeter amplitude associated with the pion  
in the medium is given 
by~\cite{deMelo2014,deMelo2002}
\begin{eqnarray}
\Psi^*(k+V,P) = \frac{\rlap\slash{k}+\rlap\slash V+m_q^*}{(k+V)^2-m_q^{*2}+ i\epsilon}
\gamma^5 \Lambda^* (k+V,P)
\frac{\rlap\slash{k}+\rlap\slash V-\rlap\slash{P}+m_q^*}{(k+V-P)^2-m_q^{*2}+ i\epsilon}.
\label{bsa}
\end{eqnarray}
In the expression above, the instantaneous terms are separated out in the quark propagators, 
and $k^\mu + \delta^\mu_0 V^0~\rightarrow~k^\mu$ shift will be made for  
all the relevant places. Performing the light-front energy 
integration,~$k^-$, the valence pion wave function is obtained:  
\begin{equation}
\Phi^*(k^+,\vec k_\perp; P^+,\vec P_\perp)=
\frac{P^+}{m^{*2}_\pi-M^2_0}  
\left[\frac{N^*}
{(1-x)(m^{*2}_{\pi}-{\cal M}^2(m_q^{*2}, m_R^2))} 
 +\frac{N^*}
{x(m^{*2}_{\pi}-{\cal M}^2(m^{2}_R, m_q^{*2}))} \right].
\label{wf2}
\end{equation}
Here $N^*$ is a normalization factor,
$N^*=C^*\frac{m_q^*}{f^*_\pi}(N_c)^\frac12$, and $x=k^+/P^+$  
with \ $0 \le x \le 1$, 
${\cal M}^2(m^2_a, m_b^2)= \frac{k^2_\perp+m_a^2}{x}+\frac{%
(P-k)^2_\perp+m^2_{b}}{1-x}-P^2_\perp \ $,
and the square of the mass is $M^2_0 ={\cal M}^2(m_q^{*2}, m_q^{*2})$. 
Note that the normalization factor is also affected in the medium, and 
the condition $F^*_{\pi}(q^2=0)=1$ (the pion charge) is imposed to fix the 
normalization factor.

The final expression for the pion electromagnetic form factor in symmetric nuclear matter 
is given by:
\begin{eqnarray}
F_\pi^{*(WF)}(q^2)& = & \frac{1}{2\pi^3(P^{\prime +}+P^+)}
           \int \frac{d^{2} k_{\perp} d k^{+} 
        \theta (k^+)\theta(P^+-k^+)}{k^+(P^+-k^+) (P^{^{\prime}+}-k^+)}
         \Phi^*(k^+,\vec k_\perp;P^{\prime +},\frac{\vec q_\perp}{2}) 
         \nonumber \\
& \times & \left (k^-_\mathrm{on}P^+P^{\prime +}-
\frac{1}{2} \vec k_\perp \cdot \vec q_\perp (P^+-P^{\prime +})-
\frac{1}{4}\, k^+q^2_\perp \right  ) 
\Phi^*(k^+,\vec k_\perp;P^{ +},-\frac{\vec q_\perp}{2}) \ .
\label{Fwf}
\end{eqnarray}

Next, we calculate the probability of the 
valence~$q\bar{q}$ state for the pion~\cite{deMelo2014,deMelo2002}, 
\begin{eqnarray}
f^*(k_\perp)= \frac{1}{4\pi^3 m^*_\pi} \int_0^{2\pi} d\phi \int^{P^+}_0 \frac{d k^{+}M_0^{*2}}
{k^+(P^+-k^+)} \Phi^{*2}(k^+,\vec k_\perp;m^*_\pi,\vec 0),
\label{prob1}
\end{eqnarray}
and  the integration of $f^*(k_\perp)$, gives 
\begin{eqnarray}
\eta^*=\int^\infty_0 dk_\perp k_\perp f^*(k_\perp), 
\label{Eq:vcompo}~
\end{eqnarray}
which gives the probability of the valence component of the 
pion in symmetric nuclear matter.

In addition, we calculate the in-medium pion decay constant, $f^*_{\pi}$, 
from the axial current:
\begin{equation}
P_\mu \langle 0(\rho)|A^\mu_i |\pi^*_j \rangle  = i m_\pi^{*2} f^*_\pi \delta_{ij}
\simeq i m_\pi^2 f^*_\pi \delta_{ij}.
\end{equation}
Using~$A^\mu_i = \bar{q} \gamma^\mu \gamma^5 \frac{\tau_i}{2} q$, the 
Lagrangian density~\cite{deMelo2014}, and after the $k^-$~integration, 
the final expression for the in-medium decay constant is obtained as
\begin{eqnarray}
f^*_{\pi} = \frac{m_q^*(N_c)^\frac12}{4\pi^3} \int \frac{d^{2} k_{\perp} d k^+ } 
{k^+(P^+-k^+)}\,
\Phi^*(k^+,\vec k_\perp;m^*_\pi,\vec 0),
\label{fpi}
\end{eqnarray}
where the expression above is associated with the plus-component of the axial current.  
Thus, $f^*_\pi$ cannot be separated into the time and space components 
as done in chiral perturbation theory~\cite{Hayano,Kienle,Vogl,Kirchbach,Meissner,Goda}.

\section{Results and discussions}

\begin{figure*}[t]
\begin{center}
\includegraphics[scale=.65]{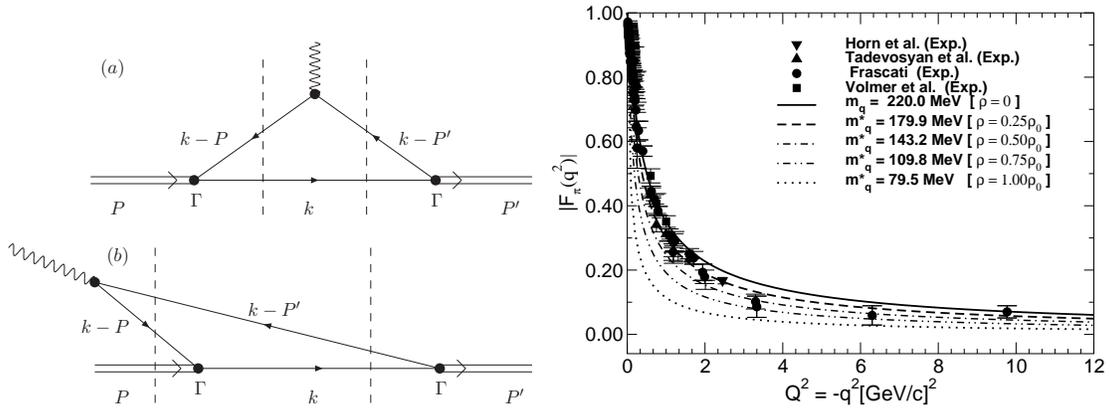}  
\includegraphics[scale=0.30]{pifofactor2.eps}
\caption{(Left)~Feymman triangle diagramm:~(a)~valence contribution, 
$F_{\pi}^{I}$, and (b), non-valence contribuitions,~$F^{II}_{\pi}$.
\newline
(Right)~Electromagnetic form factor of the pion in symmetric nuclear matter, 
calculated for several nuclear densities, compared with the experimental 
data in the vacuum from~Refs.~\cite{amen,tj,Baldini2000,cea,corn1,corn2,bebek}.
\label{Fig2}
}
\end{center}
\end{figure*}

\begin{table}[htb]
\caption{
Summary of in-medium pion properties. $\eta^*$ is calculated 
via Eq.~(\ref{prob1}), the probability of the valence quark 
component in the pion.
}
\vspace*{3mm}
\begin{tabular}{|c|c|c|c|c|}
\hline
$\rho/\rho_0$  & $m^*_q$~[MeV] & $f^*_{\pi}$~[MeV] & $<r^{*2}_{\pi}>^{1/2}$~[fm] & $\eta^*$\\
\hline
~0.00     & ~220  & ~93.1   & ~0.73   & ~0.782 \\
~0.25  &  ~179.9  & ~80.6   & ~0.84   & ~0.812 \\
~0.50  &  ~143.2  & ~68.0   & ~1.00   & ~0.843 \\
~0.75  &  ~109.8  & ~55.1   & ~1.26   & ~0.878 \\
~1.00  &  ~79.5   & ~40.2   & ~1.96   & ~0.930 \\
\hline
\end{tabular}
\label{table1}
\end{table}

The model presented here has two free parameters, i.e., the constituent quark mass  
$m_q=0.220$~GeV used to describe the pion 
properties~\cite{Fredereico92,Salme1995,Isgur1985}, 
and the regulator mass,~$m_R=0.600$~GeV, the same value used for the pion in the
vacuum~\cite{deMelo2002,Yabuzaki2015}. 
The value of $m_R$ is obtained by the fit to the experimental value of the in-vacuum 
pion decay constant~(see table~\ref{table1}), i.e., 
$f_\pi=92.4$~MeV~\cite{PDG}. In the (cold) nuclear medium, the pion mass is 
approximately given by the in-vacuum value, $m^*_\pi \simeq m_\pi=140$~MeV
(see Refs.~\cite{Hayano,Vogl,Meissner} for discussion). 
The electromagnetic radius is calculated from the derivative 
of the electromagnetic form factor at a very low 
momentum, and 
with the parameters above, the radius obtained 
in the vacuum is $r_\pi=0.74$~fm~\cite{deMelo2014,deMelo2002}, which is 
very close to the experimental value of~$0.67\pm0.02$~fm~\cite{PDG}.

In this work, we have studied the pion properties in symmetric nuclear matter 
with the light-front constituent quark model plus the QMC model.
We have calculated the pion electromagnetic form factor, electromagnetic radius, 
valence quark component probability, up to the nuclear matter density with the plus 
component of the electromagnetic current in the light-front approach.

The results show that the electromagnetic form factor decreases with increasing 
the nuclear density as can be seen in Fig.~(2b). 
Furthermore, the electromagnetic radius increases 
as the nuclear density increases~(see Table I)  
(since it depends on the in-medium electromagnetic form factor).
The in-medium pion decay constant decreases with increasing the nuclear density,  
and this agrees with the conclusion extracted from the pionic-atom experiments. 
Also the valence quark component probability in the medium,~$\eta^*$, increases with 
increasing the nuclear density as shown in Table~\ref{table1}. 
This is because the decrease of the in-medium constituent quark mass   
makes it easier to excite the valence constituent quark, and yields to 
a larger valence quark distribution inside the pion.

In the near future we plan to explore the in-medium properties of kaons and D-mesons,
as well as the vector particles like $\rho$ and $\omega$ mesons.  
Such studies are under in progress.

\vspace{0.5cm}
\noindent
{\bf Acknowledgement}\\
This work was partially supported by the Brazilian agencies 
CNPq,~FAPESP~and~Universidade Cruzeiro do Sul~(UNICSUL). 
The authors thank the organizers of {\it XVI International Conference on Hadron Spectroscopy} 
for the invitations, and hospitality at JLab during the worshop.

\end{document}